\documentclass[prb, showpacs]{revtex4}
\usepackage{bm}
\usepackage{graphicx}%
\usepackage{epstopdf}
\begin{document}
\title
{Diamagnetism  and suppression of screening as hallmarks of
electron-hole pairing in a double layer graphene system}
\author{K.\,V.\,Germash, D.\,V.\,Fil }
\affiliation{Institute for Single
Crystals, National Academy of Sciences of Ukraine, Lenin Ave. 60,
Kharkov 61001, Ukraine}

\begin{abstract}

We study  how the electron-hole pairing reveals itself in the
response of a double layer graphene system to the vector and scalar
potentials. Electron-hole pairing results in a rigid (London)
relation between the current and  the difference of vector
potentials in two adjacent layers. The diamagnetic effect can be
observed in multiple connected systems in the magnetic field
parallel to the graphene layers. Such an effect would be considered
as a hallmark of the electron-hole pairing, but the value of the
effect is extremely small. Electron-hole pairing significantly changes 
the response to the scalar potential, as well. It
results in a complete (at zero temperature) or partial (at finite
temperature) suppression of screening of the electric field of a
test charge situated at some distance to the double layer system. A
strong increase of the electric field induced by the test charge
under decrease in temperature can be considered as a spectacular
hallmark of the electron-hole pairing.
\end{abstract}


\maketitle

\section{Introduction}

The idea to realize the electron-hole pairing and the counterflow
superconductivity in adjacent electron-doped and hole-doped
conducting layers  was put forward in Refs. \onlinecite{1,2}.
Electron-hole pairs with spatially separated components may provide
a flow of equal in modulus and oppositely directed electrical
currents in the
 double layer system. The transition of the gas
of such pairs into the superfluid state can be considered as a kind
of the superconductive transition.

The electron-hole pairing was also predicted for the quantum Hall
bilayers \cite{3,3a,3b} with the total filling factor of Landau
levels $\nu_1+\nu_2=1$ ($\nu_i$ is the filling factor of the $i$th
layer). In the latter systems  empty states in the zeroth Landau
level play the role of holes. The electron-hole pairing in  quantum
Hall bilayers was observed experimentally. The pairing results in
the exponential increase of the counterflow conductivity \cite{5, 5a,
5b}, and the negative perfect interlayer drag \cite{6} at low
($T<0.1$ K) temperature.

The idea to realize electron-hole pairing in a double layer graphene
system  was put forward in Refs. \onlinecite{g1,g2,g3}. In graphene
the electron and hole Fermi circles are nested at equal densities of
carriers, which is the favorable condition for the electron-hole
pairing. In addition, the type of carriers (electrons or holes) and
their densities can be easily controlled by external gates. The
estimate for the critical temperature given in Ref. \onlinecite{g2}
is very optimistic ($T_c\approx 300$ K).

It was shown \cite{khe} that screening may significantly (in six
orders) reduce the critical temperature. The pessimistic
conclusion \cite{khe} was put in question in Refs.
\onlinecite{g4,g5}. The electron-hole pairing suppresses screening
and the problem should be treated self-consistently. If the bare
coupling constant is quite large, the influence of screening on the
critical temperature is not so crucial. Using a self-consistent
treatment of dynamic screening, the authors of Ref. \onlinecite{g4}
have shown that at sufficiently small interlayer distance the
excitonic gap (the order parameter for the electron-hole pairing)
can reach values of the order of the Fermi energy. In that case the
critical temperature can be rather large (up to 100 K).

It was predicted \cite{g6} that the counterflow superconductivity at
high temperature can also be observed in a pair of adjacent bilayer
graphene sheets. The advantage of the latter system is that the
coupling parameter depends on the density of carriers and this
parameter can be controlled by external gates. Using few-layer ABC
stacked sheets of graphene instead of monolayer or bilayer graphene
sheets one may further increase the critical temperature \cite{ml}.
The increasing is caused by the enhancing of the density of states
in the few-layer graphene.

Drag experiments  support the idea of the electron-hole pairing in
electron-hole  double layer systems in zero magnetic field. An
anomalous increase in the drag resistance at low temperature was
observed in GaAs-AlGaAs double quantum well heterostructures at
small ($<20$ nm) interlayer distances \cite{ex1, ex2}. The effect is
larger in samples with lower density of the carrier. The anomalous
drag was registered in a hybrid double layer system comprising a
single-layer (or bilayer) graphene in close proximity to a quantum
well created in GaAs \cite{gam}. In the latter system the value of
the anomalous drag is larger than in the GaAs double layer
structures \cite{ex1, ex2} and the effect is registered at higher
temperature. At the same time, the double layer graphene system
\cite{gex} does not show any low temperature increase in  the drag
resistance down to the temperature about 1 K. The anomalous drag can
be explained by the electron-hole pairing.  The experimental
situation correlates to the understanding \cite{g4,g5,g6} that the
critical temperature is very sensitive to the parameters of the
double layer system (the electron spectrum, the interlayer distance,
the density of the carriers, etc.).

Generally, the transition into the superconductive state is
registered not only in transport measurements, but in magnetic
measurements, as well. The Meissner effect is a known indicator of
the superconductive transition. This indicator is especially
important in a situation where only a partial lowering of the
resistance is observed  (when only part of the sample is
superconductive). The Meissner effect can be derived from the London
equation. In the case of the counterflow superconductivity the
London equation (its analog) determines the rigid relation between
the counterflow current and the difference of the vector potentials
in graphene sheets. The system demonstrates the diamagnetic response
to the magnetic field directed parallel to the layers. The effect
can be observed in a multiple-connected geometry. The density of the
induced current can be rather large, but the diamagnetic effect is
extremely small.

For the electron-hole pairing the counterpart of the Meissner effect
is the suppression of screening. The response to the difference of
the vector potentials in the layers is similar to the response to
the sum of the scalar potentials.  The suppression of screening is
specific for the counterflow superconductivity. Screening is
connected with the appearance of the induced charges in the
conducting layers. In the case of the electron-hole pairing the
positive and negative induced charges are correlated, which reduces
screening. We propose to observe the suppression of screening by the
measurement of the electrostatic field of a test charge located near
the graphene sheets. We show that at zero temperature screening
should be completely suppressed at large distance to the test
charge. At finite temperatures much lower than the critical
temperature the suppression is strong but partial. This behavior can
be considered as a spectacular hallmark of the electron-hole
pairing.

\section{Density and current operators for the monolayer and bilayer graphene}
\label{s1}

For computing the density-density and the current-current response
functions we need the explicit expressions for the density operator
and the current operator. For graphene these operators differ from 
those for a free electron gas with the quadratic spectrum. In this
section we derive the expressions for the Fourier component of the
density and the current operator for the monolayer and the bilayer
graphene.

The Hamiltonian that describes low-energy electrons in the monolayer
 and bilayer graphene has the form \cite{gr}
\begin{eqnarray} \label{1}
  H_m &=& \hbar v_F
\left(
  \begin{array}{cccc}
    0 & \hat{k}_x-\mathrm{i} \hat{k}_y & 0 & 0 \\
    \hat{k}_x+\mathrm{i} \hat{k}_y & 0 & 0 & 0 \\
    0 & 0 & 0 & \hat{k}_x+\mathrm{i} \hat{k}_y \\
    0 & 0 & \hat{k}_x-\mathrm{i} \hat{k}_y  & 0 \\
  \end{array}
\right)\otimes \mathbf{1}_{\sigma},
   \cr
  H_b &=& -\frac{\hbar^2}{2m}
\left(
  \begin{array}{cccc}
    0 & (\hat{k}_x-\mathrm{i}\hat{k}_y)^2 & 0 & 0 \\
    (\hat{k}_x+\mathrm{i} \hat{k}_y)^2 & 0 & 0 & 0 \\
    0 & 0 & 0 & (\hat{k}_x+\mathrm{i} \hat{k}_y)^2  \\
    0 & 0 & (\hat{k}_x-\mathrm{i} \hat{k}_y)^2  & 0 \\
  \end{array}
\right)\otimes \mathbf{1}_{\sigma},
              \end{eqnarray}
where $\hat{\mathbf{k}}=-\mathrm{i}\nabla $  is the wave number
operator, $\mathbf{1}_{\sigma}$ is unit matrix that acts in the spin
space, $v_F$ is the Fermi velocity for the monolayer graphene and
$m$ is the effective mass for the bilayer graphene.
 The two latter
quantities are the material parameters: $v_F \approx 10^8$ cm/s, and
$m \approx 0.03 m_e$, where $m_e$ is the free electron mass. Here and
below the index $m(b)$ corresponds to the monolayer (bilayer)
graphene.

The Hamiltonian (\ref{1}) acts in the space of the eight-component
vectors
\begin{equation}\label{2}
    \mathbf{\Phi}=\left(
           \begin{array}{c}
            \mathbf{ \Psi}_{K\uparrow} \\
            \mathbf{ \Psi}_{K'\uparrow} \\
 \mathbf{\Psi}_{K\downarrow} \\
             \mathbf{\Psi}_{K'\downarrow} \\
           \end{array}
         \right),
\end{equation}
where
\begin{equation}\label{3}
    \mathbf{\Psi}_{K\sigma}=
     \left(
           \begin{array}{c}
             \Psi_{A} \\
             \Psi_{B} \\
                        \end{array}
         \right)
\end{equation}
is the pseudospinor whose components correspond to different
graphene sublattices (labeled as $A$ and $B$),    $K$($K'$) is the
valley index, and $\sigma$ is the spin index.

Since the Hamiltonian (\ref{1}) is diagonal in the valley and spin
indexes, the  pure  spin-valley states described by the spinor
(\ref{3}) are the eigenvectors of (\ref{1}). For the pure state the
vector (\ref{2}) contains only one nonzero component
$\mathbf{\Psi}_{K\sigma}$. The eigenvalues and the eigenfunctions
for the pure states are
\begin{eqnarray}
\label{17}
  \varepsilon^{m}_{\lambda,\mathbf{k}}=\lambda \hbar v_F k,
  \quad  \mathbf{\Psi}^{(m)}_{\alpha,\sigma,\lambda,\mathbf{k}}(\mathbf{r})
  =\frac{e^{\mathrm{i}\mathbf{k}\mathbf{r}}}{\sqrt{2S}}\left(
                             \begin{array}{c}
                          e^{-\mathrm{i}\alpha\frac{\theta_\mathbf{k}}{2}} \\
                                           \lambda  e^{\mathrm{i}\alpha\frac{\theta_\mathbf{k}}{2}}\\
                                                      \end{array}
                                    \right),
   \\ \label{18}
\varepsilon^{b}_{\lambda,\mathbf{k}}=\lambda \frac{\hbar^2 k^2}{2
m_b},\quad
\mathbf{\Psi}^{(b)}_{\alpha,\sigma,\lambda,\mathbf{k}}(\mathbf{r})
=\frac{e^{\mathrm{i}\mathbf{k}\mathbf{r}}}{\sqrt{2S}}\left(
                             \begin{array}{c}
                          e^{-\mathrm{i}\alpha{\theta_\mathbf{k}}} \\
                                           -\lambda  e^{\mathrm{i}\alpha{\theta_\mathbf{k}}}\\
                                                      \end{array}
                                    \right),
\end{eqnarray}
where $\alpha=\pm 1$ corresponds to the $K (K')$ valley, and
$\lambda=\pm 1$ corresponds to the conductive (valence) subband. In
(\ref{17}) and (\ref{18}) $\theta_{\mathbf{k}}$ is the angle between
the vector $\mathbf{k}$ and the $x$ axis, and $S$ is the area of the
system. We note that due to the spin-valley degeneracy the
superposition of four pure states with the same $\mathbf{k}$ and
$\lambda$ is also the eigenvector of (\ref{1}).

 The operator of the creation (annihilation) of an electron at
the coordinate $\mathbf{r}$ can be expressed through the operators
of the creation (annihilation) of an electron in the eigenstate with
given $\lambda$ and $\mathbf{k}$:
\begin{equation}\label{4}
\Psi^+(\mathbf{r})=\sum_{\alpha,\sigma,\lambda,\mathbf{k}}\Phi_{\alpha,\sigma,\lambda,\mathbf{k}}^+(\mathbf{r})
c^+_{\alpha,\sigma,\lambda,\mathbf{k}},\quad
\Psi(\mathbf{r})=\sum_{\alpha,\sigma,\lambda,\mathbf{k}}\Phi_{\alpha,\sigma,\lambda,\mathbf{k}}(\mathbf{r})
c_{\alpha,\sigma,\lambda,\mathbf{k}}.
\end{equation}

In the second quantization representation the Hamiltonian (\ref{1})
 is diagonal in
$c_{\alpha,\sigma,\lambda,\mathbf{k}}$ operators:
\begin{equation}\label{1a}
    H=\sum_{\sigma,\alpha,\lambda,\mathbf{k}}
    \varepsilon_{\lambda,\mathbf{k}}
    c_{\sigma,\alpha,\lambda,\mathbf{k}}^+c_{\sigma,\alpha,\lambda,\mathbf{k}}.
    \end{equation}

From (\ref{4}) we obtain the Fourier component of the density
operator for the monolayer (bilayer) graphene:
\begin{eqnarray}\label{19}
    \rho^{m(b)}(\mathbf{q})=\sum_{\alpha,\sigma,\lambda,\mathbf{k},\alpha',\sigma',\lambda',\mathbf{k}'}
    \int d \mathbf{r} e^{i\mathbf{q}\mathbf{r}}\Phi^+_{\alpha',\sigma',\lambda',\mathbf{k}'}(\mathbf{r})
    \Phi_{\alpha,\sigma,\lambda,\mathbf{k}}(\mathbf{r})c^+_{\alpha',\sigma',\lambda',\mathbf{k}'}
c_{\alpha,\sigma,\lambda,\mathbf{k}}\cr
=\sum_{\alpha,\sigma,\lambda,\lambda',\mathbf{k}}f^{m(b)}_{\alpha,\lambda,\lambda',\mathbf{k},\mathbf{q}}
c^+_{\alpha,\sigma,\lambda',\mathbf{k}+\mathbf{q}}c_{\alpha,\sigma,\lambda,\mathbf{k}},
\end{eqnarray}
where
\begin{equation}\label{20}
    f^{m}_{\alpha,\lambda,\lambda',\mathbf{k},\mathbf{q}}=
    \frac{e^{\mathrm{i}\alpha\frac{\theta_{\mathbf{k}+\mathbf{q}}-\theta_{\mathbf{k}}}{2}}
    +\lambda\lambda'e^{-\mathrm{i}\alpha\frac{\theta_{\mathbf{k}+\mathbf{q}}-\theta_{\mathbf{k}}}{2}}}{2}
\end{equation}
and
\begin{equation}\label{21}
    f^{b}_{\alpha,\lambda,\lambda',\mathbf{k},\mathbf{q}}=
    \frac{e^{\mathrm{i}\alpha(\theta_{\mathbf{k}+\mathbf{q}}-\theta_{\mathbf{k}})}
    +\lambda\lambda'e^{-\mathrm{i}\alpha(\theta_{\mathbf{k}+\mathbf{q}}-\theta_{\mathbf{k}})}}{2}.
\end{equation}
Note that the density operator (\ref{19}) contains the diagonal as
well as off-diagonal in $\lambda$ terms. In other words, the density
operator cannot be presented as a sum of the valence band and the
conductive band density operators.

To obtain the operator of the current we replace the wave number
operator $\hat{\mathbf{k}}$ in  the Hamiltonian (\ref{1}) with the
gauge invariant operator $\hat{\mathbf{k}}-(e/\hbar c)\mathbf{A}$
($c$ is the light velocity) and consider a variation of the
Hamiltonian under the variation in the vector potential
$\mathbf{A}$.
The current operator is given by the equation
\begin{equation}\label{23}
  \hat{ \mathbf{ j}}(\mathbf{r})=-c\Psi^+(\mathbf{r})\frac{\delta H}{\delta
\mathbf{A}}\Psi(\mathbf{r}).
\end{equation}

For the monolayer graphene the Fourier component of the operator
(\ref{23}) is presented in the form
\begin{eqnarray}\label{5a}
   \hat{ \mathbf{j}}^{m}(\mathbf{q})=
e
v_F\sum_{\alpha,\sigma,\lambda,\lambda',\mathbf{k}}\mathbf{g}^m_{\alpha,\lambda,\lambda',\mathbf{k},\mathbf{q}}
c^+_{\alpha,\sigma,\lambda',\mathbf{k}+\mathbf{q}}c_{\alpha,\sigma,\lambda,\mathbf{k}},
\end{eqnarray}
where $\mathbf{g}^m=g_x^m \mathbf{i}_x+ g_y^m \mathbf{i}_y$,
$\mathbf{i}_x$ and $\mathbf{i}_y$ are the unit vectors along the $x$
and $y$ axes, and
\begin{eqnarray}
  g_x^m&=&
  \frac{\lambda e^{\mathrm{i}\alpha\frac{\theta_{\mathbf{k}+\mathbf{q}}+\theta_{\mathbf{k}}}{2}}+\lambda' e^{-\mathrm{i}\alpha\frac{\theta_{\mathbf{k}+\mathbf{q}}+\theta_{\mathbf{k}}}{2}}
    }{2},  \cr
 g_y^m&=&
  -\mathrm{i} \alpha \frac{\lambda
e^{\mathrm{i}\alpha\frac{\theta_{\mathbf{k}+\mathbf{q}}+\theta_{\mathbf{k}}}{2}}
-\lambda'
e^{-\mathrm{i}\alpha\frac{\theta_{\mathbf{k}+\mathbf{q}}+\theta_{\mathbf{k}}}{2}}
    }{2}.
\end{eqnarray}
We note that the current operator (\ref{5a}) does not contain the
diamagnetic term (the term proportional to $\mathbf{A}$), different from 
the current operator for a free electron gas.

 Repeating the same procedure for the bilayer graphene we obtain
\begin{eqnarray}\label{25}
   \hat{ {j}}^{b}_x(\mathbf{q})=  \frac{e \hbar}{m}
\sum_{\alpha,\sigma,\lambda,\lambda',\mathbf{k}}
\left[\left(k_x+\frac{q_x}{2} - \frac{e}{\hbar
c}A_x\right)g_x^b+\left(k_y+\frac{q_y}{2} - \frac{e}{\hbar
c}A_y\right)g_y^b\right]
c^+_{\alpha,\sigma,\lambda',\mathbf{k}+\mathbf{q}}c_{\alpha,\sigma,\lambda,\mathbf{k}},
\end{eqnarray}
\begin{eqnarray}\label{26}
    \hat{{j}}^{b}_y(\mathbf{q})=  \frac{e \hbar}{m}
\sum_{\alpha,\sigma,\lambda,\lambda',\mathbf{k}}
\left[\left(k_x+\frac{q_x}{2} - \frac{e}{\hbar
c}A_x\right)g_y^b-\left(k_y+\frac{q_y}{2} - \frac{e}{\hbar
c}A_y\right)g_x^b\right]
c^+_{\alpha,\sigma,\lambda',\mathbf{k}+\mathbf{q}}c_{\alpha,\sigma,\lambda,\mathbf{k}},
\end{eqnarray}
where
\begin{eqnarray}
  g_x^b&=&
  \frac{\lambda e^{\mathrm{i}\alpha(\theta_{\mathbf{k}+\mathbf{q}}+\theta_{\mathbf{k}})} +\lambda' e^{-\mathrm{i}\alpha(\theta_{\mathbf{k}+\mathbf{q}}+\theta_{\mathbf{k}})}
    }{2},  \cr
 g_y^b&=&
  -\mathrm{i} \alpha \frac{\lambda e^{\mathrm{i}\alpha(\theta_{\mathbf{k}+\mathbf{q}}+\theta_{\mathbf{k}})}
  -\lambda'
e^{-\mathrm{i}\alpha(\theta_{\mathbf{k}+\mathbf{q}}+\theta_{\mathbf{k}})}}{2}.
\end{eqnarray}
Formally, the operator $\hat{ \mathbf{{j}}}^{b}$  contains the
diamagnetic term, but in fact, due to the factors $g_x^b$ and
$g_y^b$ this term does not contribute to the current after the
averaging over the angle.

\section{Double layer graphene system in the Nambu representation}
\label{s2}

 Let us now consider the double layer system made of two graphene
 sheets.  The graphene sheet 1  is assumed to be the electron doped and
the graphene sheet 2, the hole doped. The density of electrons is
equal to the density of holes.

The Hamiltonian of the system has the form
\begin{equation}\label{1b}
    H=\sum_{i=1,2}\sum_{\sigma,\alpha}\sum_{\lambda,\mathbf{k}}
    \left(\varepsilon_{\lambda,\mathbf{k}}-\epsilon_{F,i}\right)
    c_{i,\sigma,\alpha,\lambda,\mathbf{k}}^+c_{i,\sigma,\alpha,\lambda,\mathbf{k}}
    +H_{int},
\end{equation}
where $i$ is the layer index, and $\epsilon_{F,1}=+\epsilon_F$, 
$\epsilon_{F,2}=-\epsilon_F$ are the Fermi energies. The Hamiltonian
$H_{int}$ includes the intralayer and interlayer Coulomb interaction
\begin{equation}\label{3a}
    H_{int}=\frac{1}{2 S}\sum_{i,j=1,2}\sum_{\mathbf{q}}V_{ij}(q)
    :\hat{\rho}_i(\mathbf{q})\hat{\rho}_j(-\mathbf{q}):,
\end{equation}
where $V_{ij}(q)$ are the Fourier components of the interaction
potential, and the notation $:\hat{\rho}\hat{\rho}:$ means the
normal ordering.

In the system with the electron-hole pairing the average $\langle
c_{1,\lambda,\mathbf{k}}^+ h^+_{2,-\lambda,-\mathbf{k}}\rangle$
(where $h^+$ is the hole creation operator in layer 2) is
nonzero. Taking into account that
$h^+_{2,\lambda,-\mathbf{k}}=c_{2,\lambda,\mathbf{k}}$ the pairing
can be equivalently described in terms of  $\langle
c_{1,\lambda,\mathbf{k}}^+ c_{2,-\lambda,\mathbf{k}}\rangle$
averages.

The mean-field Hamiltonian can be presented in the matrix form that
is the analog of the Nambu representation
\begin{equation}\label{5b}
H_{MF} =
\sum_{\sigma,\alpha}\sum_{\mathbf{k},\lambda}\hat{C}^{+}_{\lambda,\mathbf{k}}
\hat{\epsilon}_{\mathbf{\lambda,k}} \hat{C}_{\lambda,\mathbf{k}},
\end{equation}
where
\begin{equation}\label{6a}
\hat{C}_{\lambda,\mathbf{k}} =\left(
         \begin{array}{c}
        c_{1,\sigma,\alpha,\lambda,\mathbf{k}} \\
     c_{2,\sigma,\alpha,-\lambda,\mathbf{k}} \\
          \end{array}
        \right), \quad \hat{C}^+_{\lambda,\mathbf{k}}=\left(
                                                        \begin{array}{cc}
                                                           c^+_{1,\sigma,\alpha,\lambda,\mathbf{k}}
                                                            &c^+_{2,\sigma,\alpha,-\lambda,\mathbf{k}}  \\
                                                        \end{array}
                                                      \right)
 \end{equation}
and
\begin{equation}\label{7a}
\hat{\epsilon}_{\lambda,\mathbf{k}} = \xi_{\lambda,\mathbf{k}}
\hat{\tau}_3 - \Delta_{\lambda,\mathbf{k}} \hat{\tau}_1
\end{equation}
with
$\xi_{\lambda,\mathbf{k}}=\varepsilon^{m(b)}_{\lambda,\mathbf{k}}-\epsilon_F$.
Here
 $\hat{\tau}_i$ are the Pauli matrices. Since Eqs. (\ref{5b}) and (\ref{7a})
 are
the same as in the original Nambu approach \cite{na} one can use the
standard diagram technique for obtaining the response functions.

The interaction part of the Hamiltonian $H_{int}'=H-H_{MF}$
determines the self-energy contribution to the spectrum. The
condition for this contribution not to renormalize
$\Delta_{\lambda,\mathbf{k}}$ yields  the self-consistence equation
\begin{equation}\label{8a}
    \Delta_{\lambda,\mathbf{k}}=
    \sum_{\lambda',\mathbf{q}}V_{12}^{scr}(q) F^{m(b)}_{\lambda,\lambda',\mathbf{k},\mathbf{q}}
        \frac{\Delta_{\lambda',\mathbf{k}+\mathbf{q}}}{2E_{\lambda',\mathbf{k}+\mathbf{q}}}
\tanh\frac{E_{\lambda',\mathbf{k}+\mathbf{q}}}{2T},
\end{equation}
where
$E_{\lambda,\mathbf{k}}=\sqrt{\xi_{\lambda,\mathbf{k}}^2+\Delta_{\lambda,\mathbf{k}}^2}$
is the energy spectrum,
\begin{equation}\label{811}
    F^{m}_{\lambda,\lambda',\mathbf{k},\mathbf{q}}=\frac{1+\lambda\lambda'
\cos(\theta_{\mathbf{k}+\mathbf{q}}-\theta_{\mathbf{k}})}{2}, \quad
F^{b}_{\lambda,\lambda',\mathbf{k},\mathbf{q}}=\frac{1+\lambda\lambda'
\cos[2(\theta_{\mathbf{k}+\mathbf{q}}-\theta_{\mathbf{k}})]}{2},
\end{equation}
and $V_{12}^{scr}(q)$ is the potential of screened interlayer
Coulomb interaction. For obtaining this potential one can use the
random phase approximation. In the general case this approximation
yields the frequency dependent potential $V_{12}^{scr}(q,\omega)$.
In (\ref{8a}) $V_{12}^{scr}(q)=V_{12}^{scr}(q,0)$ (the static
screening approximation). This approximation overestimates the
influence of screening on the electron-hole pairing \cite{g4}. Using
the dynamical screening approximation for the system of two
suspended monolayer graphene sheets the authors of Ref.
\onlinecite{g4} have shown that the self-consistence equation has
the solution $\Delta_{\lambda,\mathbf{k}}\sim \epsilon_F$ if the
interlayer distance is small enough. Applying the dynamical
screening approach to the system of two suspended bilayer graphene
sheets we arrive at the same conclusion. We will not present here
the details. Our starting point is that under the appropriate
conditions the electron-hole pairing takes place in a system of two
monolayer or two bilayer graphene sheets at rather high temperatures
and in this case the order parameter of the electron-hole pairing
$\Delta_{\lambda,\mathbf{k}}$ is comparable in value with the Fermi
energy.

In the Nambu representation the density operators can be presented
in the following compact form:
\begin{eqnarray}
\label{p1}
 \hat{ \rho}_+(\mathbf{q}) &=& \hat{\rho}_1(\mathbf{q})+\hat{\rho}_2(\mathbf{q})=
  \sum_{\alpha,\sigma,\lambda,\lambda',\mathbf{k}}f^{m(b)}_{\alpha,\lambda,\lambda',\mathbf{k},\mathbf{q}}
\hat{C}^{+}_{\lambda',\mathbf{k}+\mathbf{q}} \tau_0
\hat{C}_{\lambda,\mathbf{k}}, \\  \label{p2}
  \hat{\rho}_-(\mathbf{q}) &=& \hat{\rho}_1(\mathbf{q})-\hat{\rho}_2(\mathbf{q})=
  \sum_{\alpha,\sigma,\lambda,\lambda',\mathbf{k}}f^{m(b)}_{\alpha,\lambda,\lambda',\mathbf{k},\mathbf{q}}
\hat{C}^{+}_{\lambda',\mathbf{k}+\mathbf{q}} \tau_3
\hat{C}_{\lambda',\mathbf{k}},
\end{eqnarray}
where $\tau_0$ is the identity matrix.

The current operators for the system of two  monolayer graphenes can be written as
\begin{eqnarray}
\label{p3}
 \hat{ \mathbf{j}}_+(\mathbf{q}) &=& \hat{\mathbf{j}}_1(\mathbf{q})+\hat{\mathbf{j}}_2(\mathbf{q})=
  ev_F\sum_{\alpha,\sigma,\lambda,\lambda',\mathbf{k}}
  \mathbf{g}^{m}_{\alpha,\lambda,\lambda',\mathbf{k},\mathbf{q}}
\hat{C}^{+}_{\lambda',\mathbf{k}+\mathbf{q}} \tau_3
\hat{C}_{\lambda,\mathbf{k}}, \\ \label{p4}
  \mathbf{j}_-(\mathbf{q}) &=&\hat{\mathbf{j}}_1(\mathbf{q})-\hat{\mathbf{j}}_2(\mathbf{q})=
  e v_F\sum_{\alpha,\sigma,\lambda,\lambda',\mathbf{k}}\mathbf{g}^{m}_{\alpha,\lambda,\lambda',\mathbf{k},\mathbf{q}}
\hat{C}^{+}_{\lambda',\mathbf{k}+\mathbf{q}} \tau_0
\hat{C}_{\lambda,\mathbf{k}}.
\end{eqnarray}

The important feature of (\ref{p3}),(\ref{p4}) and (\ref{p1}),(\ref{p2}) 
is that the operator $\hat{ \rho}_+$ contains the same
matrix ($\tau_0$) as the  operator $\hat{ \mathbf{j}}_-$. It
reflects the connection between the Meissner effect and the
suppression of screening. The Meissner effect is determined by the
behavior of the $\mathbf{j}_-$ response function. In its turn, the
screening at large distances is determined by the behavior of the
$\rho_+$ response function.

Here we do not present the expression for $\hat{ \mathbf{j}}_\pm$
for the system of two bilayer graphenes. These expressions are just
the straightforward generalization of (\ref{25}),(\ref{26}) and
(\ref{p3}),(\ref{p4}).

\section{London equation for the counterflow superconductor}
\label{s3}

Phenomenologically,  the electron-hole pair can be described as a
polar particle with the  mass  equal to the sum of the effective
electron and hole masses $M=m^*_e+m^*_h$. Such a description works
well for the double layer system with the quadratic spectrum of
carriers \cite{bal}. The superfluid velocity $\mathbf{v}_s$ for the
Bose-Einstein condensate (quasicondensate) of such pairs  is
proportional to the gradient of the phase of the condensate wave
function: $\mathbf{v}_s=(\hbar/M)\nabla\varphi$. The electric
currents connected with the flow of electron-hole pairs are
\begin{equation}\label{51}
   \mathbf{j}_1=-\mathbf{j}_2=e n_s \mathbf{v}_s=\frac{e\hbar n_s
}{M}\nabla\varphi,
\end{equation}
where $n_s$ is the superfluid density. At $T=0$ the superfluid
density is equal to the electron (hole) density.

In the magnetic field the polar particles feel the effective vector
potential proportional to the vector product of the magnetic field
and the dipole moment of the particle $\mathbf{p}$:
$\mathbf{A}_{eff}=\mathbf{B}\times\mathbf{p}/e$ \cite{sh1,son}. For
the planar system
$\mathbf{A}_{eff}=\mathbf{A}_{1,pl}-\mathbf{A}_{2,pl}$, where
$\mathbf{A}_{i,pl}$ is the planar component of the vector potential
in the $i$th plane. In this case the electric current is given by
the equation
\begin{equation}\label{41}
    \mathbf{j}_1=-\mathbf{j}_2=\frac{e\hbar n_s
    }{M}\left[\nabla\varphi-\frac{e}{\hbar
c}\left(\mathbf{A}_{1,pl}-\mathbf{A}_{2,pl}\right)\right].
\end{equation}

Superfluid density is connected with another important quantity -
the superfluid stiffness $\rho_s$. This quantity determines the
temperature of the superfluid transition $T_c$ in a two-dimensional
nonideal Bose gas. The temperature $T_c$ satisfies the equation
\cite{kt}
\begin{equation}\label{41a}
    T_c=\frac{\pi}{2}\rho_s(T_c)
\end{equation}
($\rho_s$ depends on the temperature). Superfluid stiffness is
defined as the coefficient of expansion of the free energy in the
phase gradient
\begin{equation}\label{42}
    F=F_0+\frac{1}{2}\int d^2 r \rho_s (\nabla\varphi)^2.
\end{equation}

In the presence of the vector potential the phase gradient should be
replaced with the gauge-invariant quantity
$\nabla\varphi-\frac{e}{\hbar c}\mathbf{A}_{eff}$. The variation of
the free energy with respect to the vector potential yields
\begin{equation}\label{43}
\delta F=-\int d^2 r\frac{e}{\hbar
c}\rho_s\left[\nabla\varphi-\frac{e}{\hbar
c}\left(\mathbf{A}_{1,pl}-\mathbf{A}_{2,pl}\right)\right](\delta\mathbf{A}_{1,pl}-\delta\mathbf{A}_{2,pl}).
\end{equation}
On the other hand
\begin{equation}\label{43a}
\delta F=-\frac{1}{c}\int d^2 r \left( \mathbf{j}_1 \delta
\mathbf{A}_{1,pl}+\mathbf{j}_2 \delta \mathbf{A}_{2,pl}\right).
\end{equation}
As follows from Eqs. (\ref{43}) and (\ref{43a}), the superfluid
stiffness   is the coefficient of proportionality between the
electric current and the gauge-invariant quantity $(e/\hbar )[\nabla
\varphi -(e/\hbar c)\mathbf{A}_{eff}]$. In particular, at
$\nabla\varphi=0$ we obtain
\begin{equation}\label{45}
 \mathbf{j}_-=\mathbf{j}_1-\mathbf{j}_2=-\frac{2 e^2 \rho_s}{\hbar^2 c}
\left(\mathbf{A}_{1,pl}-\mathbf{A}_{2,pl}\right).
\end{equation}
Equation (\ref{45}) can be considered as an analog of the London
equation for the counterflow superconductor.

The phenomenological approach \cite{bal} yields the following expression
for the superfluid stiffness:
\begin{equation}\label{44}
   \rho_s=\frac{\hbar^2 n_s}{M}.
\end{equation}
The same expression appears in the microscopic theory of the
electron-hole pairing \cite{1}.

Since the mass of the carrier in the monolayer graphene is equal to
zero, Eq. (\ref{44}) cannot be applied to the system of two
monolayer graphenes.  To overcome this difficulty we take into
account that the current can be found  as a linear response  to the
vector potential. The superfluid stiffness can be obtained as the
corresponding limit of the current-current response function.

The linear response theory yields the following equation for the
electric current induced by the vector potential:
\begin{equation}\label{46}
j_{\pm,\mu}(\mathbf{q})=-\Pi_{\pm,\mu\nu}(\mathbf{q})A_{\pm,\nu}(\mathbf{q}),
\end{equation}
where $A_{\pm,\nu}(\mathbf{q})=A_{1,\nu}(\mathbf{q})\pm
A_{2,\nu}(\mathbf{q})$, and
\begin{equation}\label{47}
    \Pi_{\pm,\mu\nu}(\mathbf{q})=-\frac{1}{2c S}
    \int_0^\beta d \tau \langle T_\tau \hat{j}_{\pm,\mu}(\mathbf{q},\tau)
    \hat{j}_{\pm,\nu}(-\mathbf{q},0)\rangle
\end{equation}
is the current-current response function. In (\ref{47}) $\beta=1/T$,
$T_\tau$ is the imaginary time ordering operator, and
$$
\hat{j}_{\pm,\mu}(\mathbf{q},\tau)=e^{H_{MF}\tau}\hat{j}_{\pm,\mu}(\mathbf{q})e^{-H_{MF}\tau}.$$
As was already mentioned in Sec. \ref{s2} the current operator does
not contain the diamagnetic term, different from the case
considered in Ref. \onlinecite{1}. We emphasize that it does not
mean the absence of the diamagnetic effect. The diamagnetic response
is included into the current-current response function.

 The response functions (\ref{47})
are similar to those that appear in the Bardeen-Cooper-Schrieffer
(BCS) theory of superconductivity \cite{bcs}. For the system of two
monolayer graphenes it is equal to
\begin{equation}\label{48}
\Pi_{\pm,\mu\nu}(\mathbf{q})=-\frac{e^2
v_F^2}{c}\frac{\delta_{\mu\nu}}{S}\sum_{\sigma,\alpha,\lambda,\lambda',\mathbf{k}}
F_{\nu,\lambda,\lambda',\mathbf{k},\mathbf{q}}
\left[P^\pm_{\lambda,\lambda',\mathbf{k},\mathbf{q}}\frac{1-f_F(E_{\lambda',\mathbf{k}+\mathbf{q}})-
f_F(E_{\lambda,\mathbf{k}})}{E_{\lambda',\mathbf{k}+\mathbf{q}}+E_{\lambda,\mathbf{k}}}+
L^\pm_{\lambda,\lambda',\mathbf{k},\mathbf{q}}\frac{f_F(E_{\lambda',\mathbf{k}+\mathbf{q}})-
f_F(E_{\lambda,\mathbf{k}})}{E_{\lambda,\mathbf{k}}-E_{\lambda',\mathbf{k}+\mathbf{q}}}
\right],
\end{equation}
where
\begin{equation}\label{49}
    F_{x,\lambda,\lambda',\mathbf{k},\mathbf{q}}=\frac{1+
    \lambda\lambda'\cos(\theta_{\mathbf{k}+\mathbf{q}}+\theta_{\mathbf{k}})}{2},
    \quad  F_{y,\lambda,\lambda',\mathbf{k},\mathbf{q}}=\frac{1-
    \lambda\lambda'\cos(\theta_{\mathbf{k}+\mathbf{q}}+\theta_{\mathbf{k}})}{2}
\end{equation}
is the factor caused by the chirality of the graphene wave function,
\begin{equation}\label{50}
P^\pm_{\lambda,\lambda',\mathbf{k},\mathbf{q}}=\frac{1}{2}
\left(1-\frac{\xi_{\lambda',\mathbf{k}+\mathbf{q}}\xi_{\lambda,\mathbf{k}}\mp
\Delta_{\lambda',\mathbf{k}+\mathbf{q}}\Delta_{\lambda,\mathbf{k}}}
{E_{\lambda',\mathbf{k}+\mathbf{q}}E_{\lambda,\mathbf{k}}}\right),
\end{equation}
\begin{equation}\label{51c}
L^\pm_{\lambda,\lambda',\mathbf{k},\mathbf{q}}=\frac{1}{2}
\left(1+\frac{\xi_{\lambda',\mathbf{k}+\mathbf{q}}\xi_{\lambda,\mathbf{k}}\mp
\Delta_{\lambda',\mathbf{k}+\mathbf{q}}\Delta_{\lambda,\mathbf{k}}}
{E_{\lambda',\mathbf{k}+\mathbf{q}}E_{\lambda,\mathbf{k}}}\right)
\end{equation}
are familiar in the BCS theory \cite{bcs} coherence factors, and
\begin{equation}\label{52}
f_F(E)=\frac{1}{\exp(E/T)+1}
\end{equation}
is the Fermi distribution function.

We will analyze the expression (\ref{48}) in the constant gap
approximation ($\Delta_{\lambda,\mathbf{k}}=\Delta=const$) at $T=0$.
To obtain the London equation from (\ref{46}) and (\ref{48}) we
consider the $\mathbf{q}\to 0$ limit. Then the integral in
(\ref{48}) can be computed analytically:
\begin{eqnarray}
\label{54} \lim_{\mathbf{q}\to
0}\Pi_{+,xx}(\mathbf{q})=-\frac{e^2}{2\pi\hbar^2 c}\epsilon_m\Bigg[
\frac{\tilde{\epsilon}_m+1}{\sqrt{(\tilde{\epsilon}_m+1)^2+\tilde{\Delta}^2}}+
\frac{\tilde{\epsilon}_m-1}{\sqrt{(\tilde{\epsilon}_m-1)^2+\tilde{\Delta}^2}}\Bigg],
\end{eqnarray}
\begin{eqnarray}\label{55}
\lim_{\mathbf{q}\to 0}\Pi_{-,xx}(\mathbf{q})=-\frac{e^2}{2\pi\hbar^2
c}\epsilon_F\Bigg[\sqrt{(\tilde{\epsilon}_m+1)^2+\tilde{\Delta}^2}+
\sqrt{(\tilde{\epsilon}_m-1)^2+\tilde{\Delta}^2}\cr
+\tilde{\Delta}^2\ln
\left(\frac{\sqrt{(\tilde{\epsilon}_m+1)^2+\tilde{\Delta}^2}+\tilde{\epsilon}_m+1}
{\sqrt{(\tilde{\epsilon}_m-1)^2+\tilde{\Delta}^2}+\tilde{\epsilon}_m-1}\right)
-2\sqrt{1+\tilde{\Delta}^2}-\tilde{\Delta}^2\ln
\left(\frac{\sqrt{1+\tilde{\Delta}^2}+1}{\sqrt{1+\tilde{\Delta}^2}-1}\right)\Bigg].
\end{eqnarray}
Here $\epsilon_m=\hbar v_F k_m$, where $k_m$ is the ultraviolet wave
vector cutoff, $\tilde{\epsilon}_m=\epsilon_m/\epsilon_F$, and
$\tilde{\Delta}=\Delta/\epsilon_F$. The quantities (\ref{54}),(\ref{55}) 
depend on the cut-off value $k_m$. That is the known
shortage \cite{vec} of the linear approximation for the spectrum.
This approximation is not valid in the  whole Brillouin zone, while
the correct answer can be obtained by the integration over the whole
Brillouin zone.

The regularization of the answer (\ref{54}),(\ref{55}) is based on
the requirement of the absence of the Meissner effect in the normal
state.
 Due to noncommutativity of the limits $T\to 0$ and
$\Delta \to 0$ we should return to Eq. (\ref{48}). At $\Delta=0$ Eq.
(\ref{48}) is reduced to
\begin{equation}\label{56}
\Pi^0_{+,\mu\nu}(\mathbf{q})=\Pi^0_{-,\mu\nu}(\mathbf{q})=-\frac{e^2
v_F^2}{c}\frac{\delta_{\mu\nu}}{S}\sum_{\sigma,\alpha,\lambda,\lambda',\mathbf{k}}
F_{\nu,\lambda,\lambda',\mathbf{k},\mathbf{q}}
\frac{f_F(\xi_{\lambda',\mathbf{k}+\mathbf{q}})-
f_F(\xi_{\lambda,\mathbf{k}})}{\xi_{\lambda,\mathbf{k}}-\xi_{\lambda',\mathbf{k}+\mathbf{q}}}.
\end{equation}
From (\ref{56}) we obtain
\begin{eqnarray}
\label{57} \lim_{\mathbf{q}\to
0}\Pi^0_{+,xx}(\mathbf{q})=\lim_{\mathbf{q}\to
0}\Pi^0_{-,xx}(\mathbf{q}) =-\frac{e^2}{\pi\hbar^2 c}\epsilon_m.
\end{eqnarray}
The contribution of the rest of the Brillouin zone should compensate
the quantity (\ref{57}). The regularized response function is
obtained by extracting from (\ref{54}),(\ref{55}) the quantity
(\ref{57}) and taking the limit $\tilde{\epsilon}_m\to \infty$. The
result is
\begin{eqnarray}
\label{541} \lim_{\mathbf{q}\to 0}\Pi^{r}_{+,xx}(\mathbf{q})=0,
\quad \lim_{\mathbf{q}\to
0}\Pi^{r}_{-,xx}(\mathbf{q})=\frac{e^2}{\pi\hbar^2
c}\epsilon_F\Bigg[\sqrt{1+\tilde{\Delta}^2}+\frac{\tilde{\Delta}^2}{2}\ln
\left(\frac{\sqrt{1+\tilde{\Delta}^2}+1}{\sqrt{1+\tilde{\Delta}^2}-1}\right)\Bigg].
\end{eqnarray}

Substituting Eq. (\ref{541}) into Eq. (\ref{46}) and doing the
reverse Fourier transformation we obtain the rigid relation between
the supercurrent $\mathbf{j}_-$ and the difference of the vector
potentials $\mathbf{A}_-$ in the form (\ref{45}). As was
expected, there is no rigid relation between  $\mathbf{j}_+$ and
$\mathbf{A}_+$. The current $\mathbf{j}_+$ is not connected with the
motion of electron-hole pairs and it is not excited by the constant
magnetic field. From (\ref{541}) we see that the London rigidity
depends on $\Delta$, different from the case of bulk
superconductors, where the London rigidity at $T=0$ is determined by
the electron density and independent of the BCS order parameter.

To obtain the superfluid stiffness we should take into account that
the response function contains the contribution of four spin-valley
components. The superfluid stiffness per the component is
\begin{equation}\label{551}
    \rho^m_s= \frac{1}{4} \frac{\hbar^2 c}{2 e^2}\lim_{\mathbf{q}\to
0}\Pi^{r}_{-,xx}(\mathbf{q})=\frac{\epsilon_F}{8
\pi}\Bigg[\sqrt{1+\tilde{\Delta}^2}+\frac{\tilde{\Delta}^2}{2}\ln
\left(\frac{\sqrt{1+\tilde{\Delta}^2}+1}{\sqrt{1+\tilde{\Delta}^2}-1}\right)\Bigg].
\end{equation}
For $\Delta \ll \epsilon_F$ the superfluid stiffness
$\rho_s=\epsilon_F/8\pi$, which coincides with the result of Ref.
\onlinecite{loz2}.

For the system of two bilayer graphenes the  response functions
$\Pi_{\pm,xx}$ are given by the equation
\begin{equation}\label{561}
\Pi_{\pm,xx}(\mathbf{q})=-\frac{e^2\hbar^2}{
m^2c}\frac{1}{S}\sum_{\sigma,\alpha,\lambda,\lambda',\mathbf{k}}
G^{xx}_{\lambda,\lambda',\mathbf{k},\mathbf{q}}
\left[P^\pm_{\lambda,\lambda',\mathbf{k},\mathbf{q}}\frac{1-f_F(E_{\lambda',\mathbf{k}+\mathbf{q}})-
f_F(E_{\lambda,\mathbf{k}})}{E_{\lambda',\mathbf{k}+\mathbf{q}}+E_{\lambda,\mathbf{k}}}+
L^\pm_{\lambda,\lambda',\mathbf{k},\mathbf{q}}\frac{f_F(E_{\lambda',\mathbf{k}+\mathbf{q}})-
f_F(E_{\lambda,\mathbf{k}})}{E_{\lambda,\mathbf{k}}-E_{\lambda',\mathbf{k}+\mathbf{q}}}
\right],
\end{equation}
where
\begin{eqnarray}\label{571}
    G^{xx}_{\lambda,\lambda',\mathbf{k},\mathbf{q}}=\left(k_x+\frac{q_x}{2}\right)^2
    \frac{1+\lambda\lambda'\cos[2(\theta_{\mathbf{k}+\mathbf{q}}+\theta_\mathbf{k})]}{2}+
\left(k_y+\frac{q_y}{2}\right)^2
    \frac{1-\lambda\lambda'\cos[2(\theta_{\mathbf{k}+\mathbf{q}}+\theta_\mathbf{k})]}{2}\cr
    +\lambda\lambda'\left(k_x+\frac{q_x}{2}\right)\left(k_y+\frac{q_y}{2}\right)
    \sin[2(\theta_{\mathbf{k}+\mathbf{q}}+\theta_\mathbf{k})].
\end{eqnarray}
The calculation yields the answer that also depends on the
ultraviolet cutoff. To regularize the answer we should extract the
quantity $\lim_{\mathbf{q}\to 0} \Pi^0_{\pm,xx}(\mathbf{q})$, where
\begin{equation}\label{572}
\Pi^0_{\pm,xx}(\mathbf{q})=-\frac{e^2\hbar^2}{
m^2c}\frac{1}{S}\sum_{\sigma,\alpha,\lambda,\lambda',\mathbf{k}}
G^{xx}_{\lambda,\lambda',\mathbf{k},\mathbf{q}}
\frac{f_F(\xi_{\lambda',\mathbf{k}+\mathbf{q}})-
f_F(\xi_{\lambda,\mathbf{k}})}{\xi_{\lambda,\mathbf{k}}-\xi_{\lambda',\mathbf{k}+\mathbf{q}}}
\end{equation}
is the response function in the normal state. It yields the response
function $\Pi_{-,xx}$ that in the limit $\mathbf{q}\to 0$ differs
from the function  (\ref{541}) by the factor of 2.  The superfluid
stiffness per the spin-valley component is
\begin{equation}\label{581}
    \rho^b_s= \frac{\epsilon_F}{4
\pi}\Bigg[\sqrt{1+\tilde{\Delta}^2}+\frac{\tilde{\Delta}^2}{2}\ln
\left(\frac{\sqrt{1+\tilde{\Delta}^2}+1}{\sqrt{1+\tilde{\Delta}^2}-1}\right)\Bigg].
\end{equation}
For $\Delta \ll \epsilon_F$ the superfluid stiffness
$\rho_s={\epsilon_F}/{4 \pi}$, which coincides with the
phenomenological expression (\ref{44}) in which $n$ is the electron
(hole) density per the component and $M=2m$.

Equations (\ref{55}) and (\ref{581}) are the main result of this section.
We find that the zero-temperature superfluid stiffness  increases
under increase in $\Delta$. In the small gap limit  this quantity is
determined entirely by the density of carriers in the conduction
band.

Let us now evaluate the value of the diamagnetic effect. The
diamagnetic susceptibility is determined by the ratio of the
magnetization, induced by the current $\mathbf{j}_-$, to the
external magnetic field. For $\mathbf{B}$  directed parallel to the
graphene layers  $A_{1,x}-A_{2,x}=B_y d$,  and
\begin{equation}\label{59}
    \chi= \frac{1}{4 \pi} \frac{2\pi j_{-,x}}{c B_y}=-  \frac{d
    }{2c}\lim_{q\to 0}\Pi^r_{-,xx}(q).
\end{equation}
It yields
\begin{equation}\label{591}
    \chi^m\approx-\frac{1}{2\pi} \left(\frac{e^2}{\hbar c}\right)^2\frac{1}{\alpha_{eff}}k_F d
   \end{equation}
for the system  of two monolayer graphenes, and
\begin{equation}\label{592}
    \chi^b\approx-\frac{1}{2\pi} \left(\frac{e^2}{\hbar c}\right)^2 k^2_F d a_B^{eff}
   \end{equation}
for the system  of two bilayer graphenes. Here
$\alpha_{eff}=e^2/\hbar v_F\approx 2.2$ is the effective fine-structure 
constant for the monolayer graphene,
$a_B^{eff}=\hbar^2/me^2\approx 1.5$ nm is the effective Bohr radius
for the bilayer graphene,  and $k_F$ is the Fermi wave vector.  In
these estimates we neglect the dependence of the London rigidity on
$\Delta/\epsilon_F$.

The diamagnetic effect can be observed in a multiple-connected
system (a double layer Corbino disk, a  double layer hollow
cylinder). The effect is maximal at zero temperature and it vanishes
in the normal state. The current induced by the external magnetic
field can be rather large. Taking, for instance, $B=0.1$ T,
$\epsilon_F=0.1$ eV, and $d=5$ nm, we obtain the density of the
current $j_-\sim 1$ A/m. At the same time the diamagnetic
susceptibility, which is proportional to the square of the fine-structure 
constant, is small ($4\pi\chi\sim 10^{-6} - 10^{-5}$).
Moreover, the magnetic field, induced by the current $j_-$, emerges
only in a narrow dielectric layer that separates two graphene
sheets. Therefore, it is a hard task to register this diamagnetic
effect  in the magnetic measurements.

\section{Suppression of screening caused by the electron-hole pairing}

\begin{figure}[h]
\center{\includegraphics[width=0.5\linewidth]{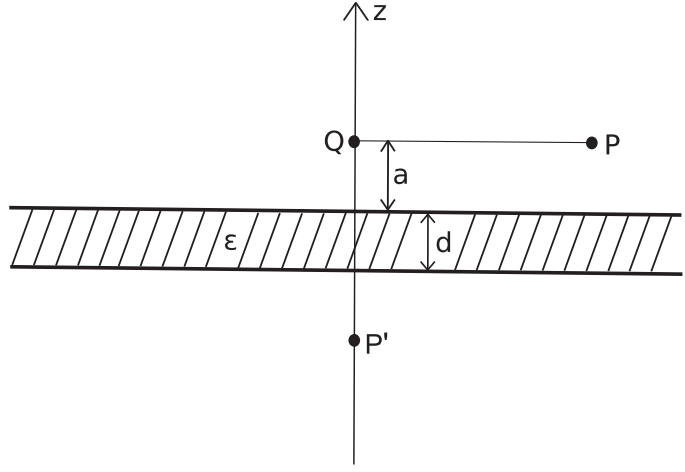}}
\caption{Schematic view of the system under consideration.}
\label{f1}
\end{figure}

Taking into account the smallness of the diamagnetic effect it is
desirable to find another effect that can be used as an indicator of
the superconductive transition. In this section we consider
screening of the electric field of a test charge. We imply that the
system (Fig. \ref{f1}) consists of two monolayer or two bilayer
graphene sheets, separated by a dielectric layer. The system is
suspended in dielectric medium with the dielectric constant close to
unity.

The electrostatic potential applied to the bilayer system results in
the appearance of the induced charges in graphene sheets
$e\rho_\pm^{ind}=e(\rho_1^{ind}\pm \rho_2^{ind})$.  They  can be
expressed through the single particle density-density response
functions:
\begin{equation}\label{60}
    e\rho^{ind}_{\pm}(\mathbf{q})=e^2\Pi_{\pm}(\mathbf{q})\varphi_\pm(\mathbf{q}),
\end{equation}
where
\begin{equation}\label{61}
    \Pi_{\pm}(\mathbf{q})=-\frac{1}{2 S}
    \int_0^\beta d \tau \langle T_\tau \hat{\rho}_{\pm}(\mathbf{q},\tau) \hat{\rho}_{\pm}(-\mathbf{q},0)
\rangle,
\end{equation}
$\varphi_{\pm}(\mathbf{q})=\varphi_1({\mathbf{q}})\pm
\varphi_2({\mathbf{q}})$, and $\varphi_i({\mathbf{q}})$ is the
two-dimensional Fourier component of the scalar potential
$\varphi(\mathbf{r})$ in the layer $i$. The potential
$\varphi(\mathbf{r})$ should be found self-consistently. It is the
sum of the bare potential created by the test charge, and the
potential caused by the induced charges. The potential
$\varphi(\mathbf{r})$ satisfies the equation
\begin{equation}\label{62}
    \nabla\left[\varepsilon(z)\nabla\varphi(\mathbf{r})\right]=-4\pi\left[Q\delta(\mathbf{r}-\mathbf{r}_Q)+
    e\rho^{ind}_1(\mathbf{r}_\perp)\delta(z-z_1)+
    e\rho^{ind}_2(\mathbf{r}_\perp)\delta(z-z_2)\right].
\end{equation}
In (\ref{62}) we consider the test charge $Q=Ze$ located at the
point $\mathbf{r}_Q=(\mathbf{r}_{Q_{\perp}},z_Q)$: the axis $z$ is
directed perpendicular to the graphene layers, $z_1$ and $z_2$ are
the coordinates of the graphene layers, and $\mathbf{r}_\perp$ is
the two-dimensional radius vector. We put $z_Q>z_1>z_2$ and
$$\varepsilon(z)=\left\{
                   \begin{array}{ll}
                     \varepsilon, & \hbox{$z_2<z<z_1$} \\
                     1, & \hbox{otherwise.}
                   \end{array}
                 \right.
$$
The solution of Eq. (\ref{62}) yields
\begin{equation}\label{63}
    e\varphi_i(\mathbf{q})=ZV_{iQ}(q)+\sum_j
V_{ij}(q)\rho^{ind}_j(\mathbf{q}),
\end{equation}
where
\begin{equation}\label{64}
    V_{11}({q})= V_{22}({q})
    =\frac{4\pi e^2}{q}\frac{(\varepsilon+1)+(\varepsilon-1)e^{-2 q d}}{(\varepsilon+1)^2-e^{-2qd}(\varepsilon-1)^2},
    \quad  V_{12}({q})= V_{21}({q})
    =\frac{8\pi e^2}{q}\frac{\varepsilon e^{-q d}}{(\varepsilon+1)^2-e^{-2qd}(\varepsilon-1)^2}
\end{equation}
are the potentials of interaction of elementary charges located in
the graphene layers,  $d=|z_1-z_2|$ is the distance between the
graphene layers, $V_{iQ}(q)=e^{-q a} V_{i1}(q)$ is the potential of
interaction between the elementary test charge and the induced
charge in the $i$th layer, and $a=|z_Q-z_1|$ is the distance from
the test charge to the nearest graphene layer. Substituting
(\ref{63}) into (\ref{60}) we obtain the induced charges
\begin{equation}\label{65}
    \rho^{ind}_{\pm}(\mathbf{q})=Ze^{-q a}
\frac{V_\pm(q)\Pi_{\pm}(\mathbf{q})}{1-V_\pm(q)\Pi_{\pm}(\mathbf{q})},
\end{equation}
where
\begin{equation}\label{651}
   V_\pm(q)=V_{11}(q)\pm V_{12}(q)=\frac{4\pi e^2}{q}\frac{1\pm
e^{-q d}}{(\varepsilon+1)\mp(\varepsilon-1)e^{- q d}}.
\end{equation}

Substituting the induced charges (\ref{65}) into Eq. (\ref{62}) and
solving it we obtain the expression for the screened potential
$\varphi(\mathbf{r})$ as the linear function of the test charge $Q$.

We consider the cases (see Fig. \ref{f1}) where the electric field
sensor is located at the side of the test charge [point P with the
coordinate
$\mathbf{r}_P=(\mathbf{r}_{Q_{\perp}}+\mathbf{r}_{\perp},z_Q)$] or
at the opposite side [point P$'$ with the coordinate
$\mathbf{r}_{P'}=(\mathbf{r}_{Q_{\perp}},z_Q-z)$, where $z>a+d$].

At the point P the potential is given by the equation
\begin{equation}\label{66}
\varphi_P(r)=\frac{Q}{r}-{Q}\int_0^{\infty}d q J_0(q r)e^{-2
qa}\left[\frac{(\varepsilon^2-1)(1-e^{-2qd})}{(\varepsilon+1)^2-e^{-2qd}(\varepsilon-1)^2}-
\frac{q}{4\pi
e^2}\left(\frac{V^2_+(q)\Pi_{+}({q})}{1-V_+(q)\Pi_{+}({q})}-
\frac{V^2_-(q)\Pi_{-}({q})}{1-V_-(q)\Pi_{-}({q})}\right) \right],
\end{equation}
where $J_0(x)$ is the Bessel function. The potential at the point
P$'$ is equal to
\begin{equation}\label{66a}
\varphi_{P'}(z)={Q}\int_0^{\infty}d q e^{-
qz}\left[\frac{4\varepsilon}{(\varepsilon+1)^2-e^{-2qd}(\varepsilon-1)^2}+
\frac{q e^{qd}}{4\pi
e^2}\left(\frac{V^2_+(q)\Pi_{+}({q})}{1-V_+(q)\Pi_{+}({q})}-
\frac{V^2_-(q)\Pi_{-}({q})}{1-V_-(q)\Pi_{-}({q})}\right) \right].
\end{equation}
In (\ref{66}) and (\ref{66a})  we take into account that
$\Pi_{\pm}(\mathbf{q})$ is the function of the modulus of the wave
vector.

 The density-density response function is  computed analogously to
the current-current response function. The answer is
\begin{equation}\label{67}
\Pi_{\pm}(\mathbf{q})=-\frac{1}{S}\sum_{\sigma,\alpha,\lambda,\lambda',\mathbf{k}}
F^{m(b)}_{\lambda,\lambda',\mathbf{k},\mathbf{q}}
\left[P^\mp_{\lambda,\lambda',\mathbf{k},\mathbf{q}}\frac{1-f_F(E_{\lambda',\mathbf{k}+\mathbf{q}})-
f_F(E_{\lambda,\mathbf{k}})}{E_{\lambda',\mathbf{k}+\mathbf{q}}+E_{\lambda,\mathbf{k}}}+
L^\mp_{\lambda,\lambda',\mathbf{k},\mathbf{q}}\frac{f_F(E_{\lambda',\mathbf{k}+\mathbf{q}})-
f_F(E_{\lambda,\mathbf{k}})}{E_{\lambda,\mathbf{k}}-E_{\lambda',\mathbf{k}+\mathbf{q}}}
\right],
\end{equation}
where  the chirality factors
$F^{m(b)}_{\lambda,\lambda',\mathbf{k},\mathbf{q}}$ are given by Eq.
(\ref{811}), and the coherence factors $P^\pm$ and $L^\pm$, by Eq.
(\ref{50}) and (\ref{51c}). The response functions (\ref{67}) do not
depend on the ultraviolet cutoff  and do not require
regularization.

We note that Eqs. (\ref{66})-(\ref{67}) correspond to the random phase
approximation (RPA) and do not account the vertex corrections. It is
known that in the BCS theory the vertex corrections are important
for obtaining the gauge-invariant result \cite{bcs}. The same is true
for the electron-hole pairing. Fortunately, the problem with the
gauge invariance does not emerge under computation of the density
response function in the static limit. Nevertheless, one can ask
about the value of the vertex corrections. This problem was
addressed in Ref. \onlinecite{nel2} where the electron-hole pairing
in bilayer systems with the quadratic spectrum of carriers (double
quantum wells is GaAs heterostructures) was studied. The good
agreement between the RPA \cite{nel2} and the diffusion quantum
Monte Carlo \cite{qmc1,qmc2} computations for the condensate
fraction allows the authors of Ref. \onlinecite{nel2} to conclude
that the vertex corrections are negligible.  The vertex corrections
for the graphene systems with electron-hole pairing were evaluated
in Ref. \onlinecite{g5}. It was shown \cite{g5} that the second-order
vertex corrections amount only to about 5\% of the first-order
coupling constants and thus can be neglected. The general argument
for neglecting the vertex corrections in the graphene system is that
they are small by the factor $1/N$, where $N$ is number of electron
flavors ($N=4$ corresponds to four spin-valley components).

\begin{figure}[h]
\center{\includegraphics[width=0.5\linewidth]{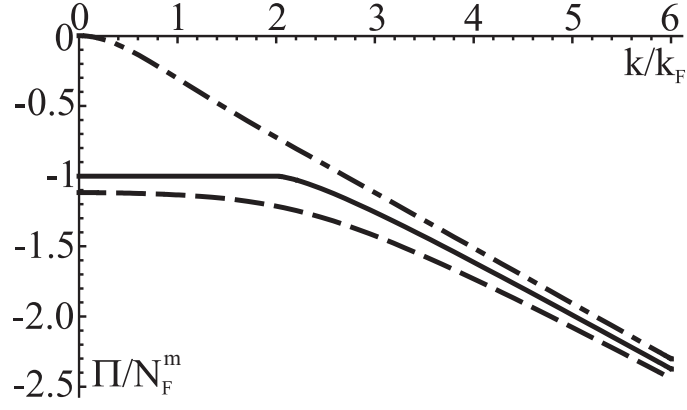}} \caption{The
density response functions $\Pi_{+}$ (dash-dotted line), $\Pi_{-}$
(dashed line) in the superconductive state (for $\Delta = 0.5
\epsilon_F$), and $\Pi=\Pi_{\pm}$ (solid line) in the normal state
for the system of two monolayer graphenes.} \label{f2}
\end{figure}

\begin{figure}[h]
\center{\includegraphics[width=0.5\linewidth]{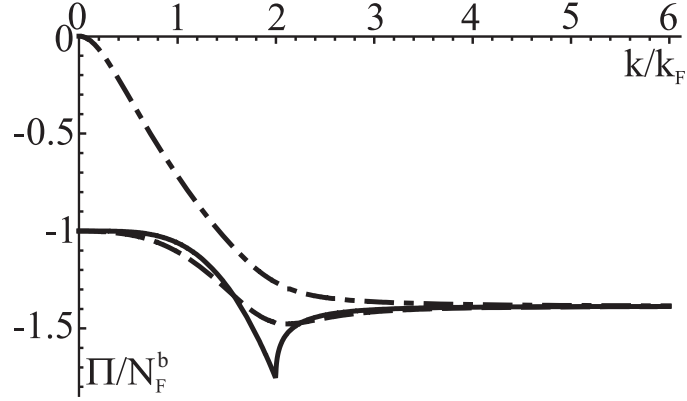}} \caption{The
same as in Fig. \ref{f2} for the system of two bilayer graphenes. }
\label{f3}
\end{figure}

Let us return to Eqs. (\ref{66})-(\ref{67}) and consider first the
$T=0$ case. The dependencies $\Pi_\pm(q)$ at $T=0$ and
$\Delta=const$ are shown in Figs. \ref{f2} and \ref{f3} for the
system of two monolayer and two bilayer graphenes, correspondingly.
 In the normal state
($\Delta=0$) the response functions $\Pi_{+}$ and $\Pi_{-}$ are
equal to each other and coincide with the response function for the
monolayer \cite{ad,sarm} (bilayer \cite{sarm1}) graphene. In the limit
$qk_F\ll 1$ they approach $\Pi_{+}(0)=\Pi_{-}(0)=N^{m(b)}_F$, the
density of states of the monolayer (bilayer) graphene at the Fermi
level ($N_F^m=2 k_F/\pi\hbar v_F$ and $N_F^b=2 m/\pi\hbar^2$). In
the superconductive state ($\Delta\ne 0$) the response function
$\Pi_+$ approaches zero at $q\to 0$. It principally changes the
character of screening.

In the normal state the scalar potential $\varphi_P(r)$ at
$d\lesssim k_F^{-1} \ll a\ll r$ has the universal asymptote
\begin{equation}\label{68}
\varphi_P(r)\approx \frac{2 Q a^2}{r^3}.
\end{equation}
Actually, it is the potential of a dipole  consisting of the test
charge and its image. In the superconductive state the second term
in (\ref{66}) yields only $1/r^3$ correction and the scalar
potential  remains unscreened at large $r$ :
$$\varphi_P(r)\approx\frac{Q}{r}.$$

Similar features demonstrates the potential $\varphi_{P'}(z)$. In
the normal state its asymptote  is
\begin{equation}\label{69}
\varphi_{P'}({z})=\frac{Q}{k_F
z^2}\frac{1}{8\alpha_{eff}\left(1+\frac{4\alpha_{eff}d
k_F}{\varepsilon}\right)}
\end{equation}
(for the system of two monolayer graphenes) and
\begin{equation}\label{70}
\varphi_{P'}({z})=\frac{Q a_B^{eff}}{ z^2}\frac{1}{8\left(1+\frac{4d
}{\varepsilon a_B^{eff}}\right)}
\end{equation}
(for the system of two bilayer graphenes). In the superconductive
state the potential remains unscreened at large $z$
[$\varphi_{P'}(z)={Q}/z$].

For $T\ne 0$, $T\ll \Delta$ the function $\Pi_+(q)$ at $q\to 0$
approaches  a small ($\propto e^{-\Delta/T}$) but finite value. It
changes the character of screening at very large distances. The
asymptotes are $\varphi_P(r)=C/r^3$, and $\varphi_{P'}(z)=C'/z^2$,
but the coefficients of proportionality $C$ and $C'$ contain the
large factor $e^{\Delta/T}$.  At $T\gg\Delta$  the screening in the
superconductive state is almost the same as in the normal state. The
screened potential at intermediate distance
 to the test charge is shown in Fig. 4 [screening along the structure, the potential $\varphi_P(r)$] and Fig. 5
 [screening across the structure, the potential $\varphi_{P'}(z)$]. The parameters
$\Delta=0.5\epsilon_F$, $d k_F = 0.2$, $a_B^{eff}k_F=0.25$, and $a
k_F=4$ are used for the computation. One can see that the
electron-hole pairing essentially changes the spatial dependence of
the screened potential both at zero and at finite temperatures.

\begin{figure}[h]
\center{\includegraphics[width=0.5\linewidth]{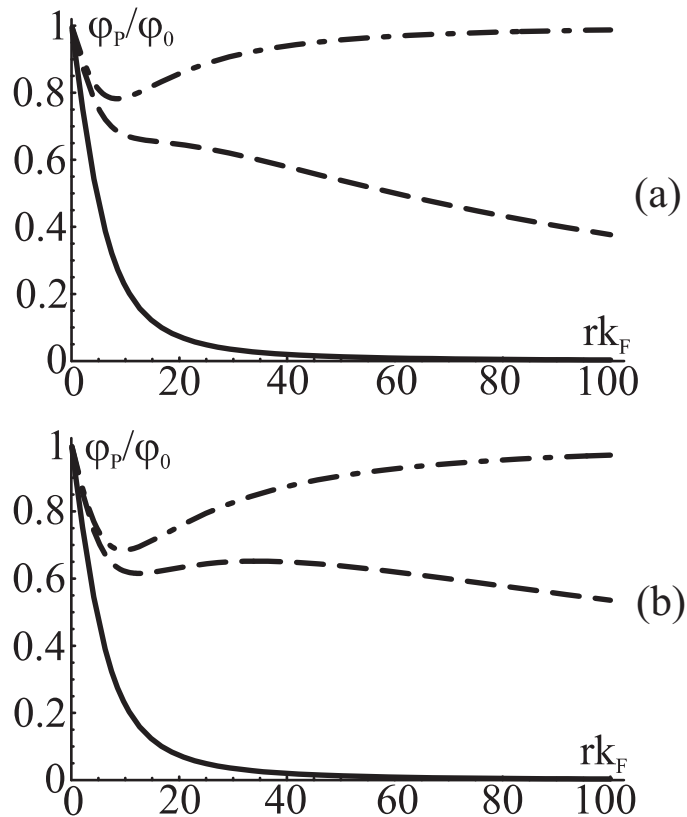}}
\caption{Screening along the structure of two monolayer (a) and two
bilayer (b) graphenes in the normal state (solid line), and in the
superconductive state at $T = 0$ (dash-dotted line) and $T=0.1\Delta$
(dashed line). The potential is normalized to the bare potential
$\varphi_0 = {Q}/{r}.$} \label{f4}
\end{figure}

\begin{figure}[h]
\center{\includegraphics[width=0.5\linewidth]{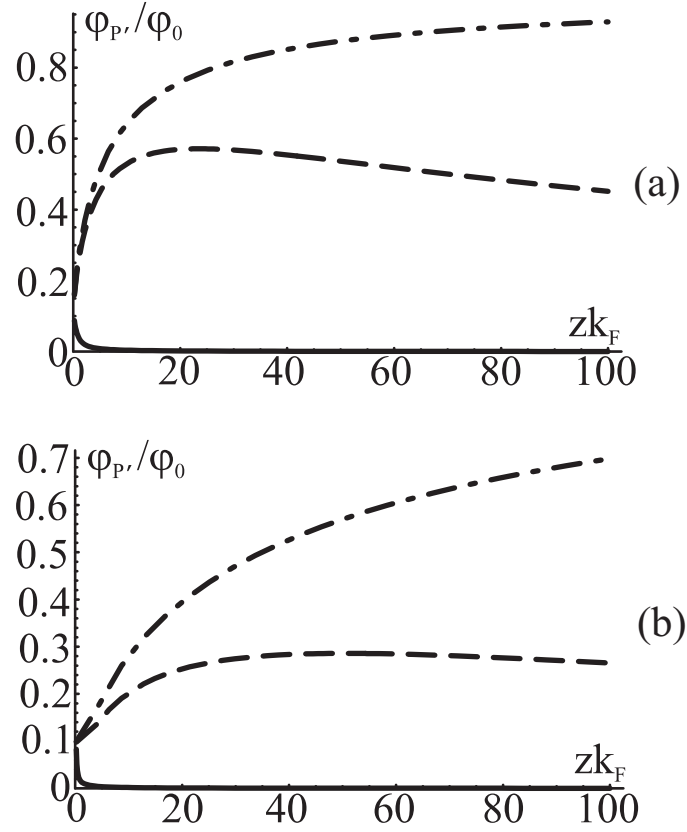}} \caption{The
same as in Fig. \ref{f4} for the screening across the structure. The
potential is normalized to $\varphi_0 = {Q}/{z}.$} \label{f5}
\end{figure}

Thus we conclude that the electron-hole pairing reveals itself in
spectacular changes of the electric field of the test charge located
near the double layer system. The effect can be used as an indicator
of the electron-hole pairing.

The most effective method of measuring the local electrostatic
potential uses the single-electron transistor (SET) technique.
Operating at low temperature the SET scanning electrometer is
capable of measuring the potential with millivolt \cite{set1} to
microvolt \cite{set2} sensitivity a high spatial resolution close to
the SET size (about 100 nm). A quantum-metrology technique for
precision three-dimensional electric-field measurement using a
single nitrogen-vacancy defect center spin in diamond was also
developed \cite{nv}. While it is less sensitive than SET, it allows
measuring the field created by an elementary charge located at a
distance less or about 150 nm from the sensor. The latter technique
does not require low temperature. Thus the electrostatic method of
registration of the electron-hole pairing is doable with the current
technologies.

\section{Conclusion}

In conclusion we have shown that in the double layer graphene system
the electron-hole pairing results in the Meissner effect and in the
strong suppression of screening of the test charge. The effects
demonstrate the same temperature behavior. They are maximal at zero
temperature, decrease under increase in the temperature, and
disappear in the normal state. It is connected with the similarity
of the current-current  and  density-density
 response functions. Such a similarity
is specific for the  electron-hole pairing.  It does not occur in
superconductors with the electron-electron pairing. The Meissner
effect in the system under study is extremely small and most
probably it cannot be used as an indicator of the electron-hole
pairing. On the other hand, the suppression of screening is strong
and the observation of this effect can be used as a hallmark of the
transition into the superconductive state.

\section*{Acknowledgment}
This work was supported by the Ukraine State Scientific and Technical Program
"Nanotechnologies and nanomaterials".

\end{document}